\documentclass[12pt]{article}
\usepackage{amsmath, amssymb, amscd, amsthm, amsfonts}
\usepackage{bm}
\usepackage{mathtools}
\usepackage{graphicx}
\usepackage{natbib}
\usepackage{url} 
\usepackage{multirow}
\usepackage{caption}
\usepackage{hyperref}
\usepackage{cite}
\usepackage{color}
\usepackage{algorithm}
\usepackage{algorithmic} 	
\usepackage{booktabs}

\makeatletter
\newenvironment{breakablealgorithm}
{
	\begin{center}
		\refstepcounter{algorithm}
		\hrule height.8pt depth0pt \kern2pt
		\renewcommand{\caption}[2][\relax]{
			{\raggedright\textbf{\ALG@name~\thealgorithm} ##2\par}%
			\ifx\relax##1\relax 
			\addcontentsline{loa}{algorithm}{\protect\numberline{\thealgorithm}##2}%
			\else 
			\addcontentsline{loa}{algorithm}{\protect\numberline{\thealgorithm}##1}%
			\fi
			\kern2pt\hrule\kern2pt
		}
	}{
		\kern2pt\hrule\relax
	\end{center}
}
\makeatother

\newcommand{\blind}{0}

\begin{document}

\def\spacingset#1{\renewcommand{\baselinestretch}%
{#1}\small\normalsize} \spacingset{1}


\if0\blind
{
  \title{\bf Bayesian Variable Selection for Single Index Logistic Model}
  \author{Yinrui Sun\\
  	School of Mathematical Sciences, Zhejiang University\\
    Department of Statistics, School of Management, Fudan University\\
    and \\
    Hangjin Jiang\thanks{Corresponding Author: H. Jiang (jianghj@zju.edu.cn); 
    	H. Jiang gratefully acknowledge The National Natural Science Foundation of China (No. 11901517).}\hspace{.2cm} \\
    Center for Data Science, Zhejiang University}
  \maketitle
} \fi

\if1\blind
{
  \bigskip
  \bigskip
  \bigskip
  \begin{center}
    {\LARGE\bf Bayesian Variable Selection for Single Index Logistic Model}
\end{center}
  \medskip
} \fi

\bigskip
\begin{abstract}
In the era of big data, variable selection is a key technology for handling high-dimensional problems with a small sample size but a large number of covariables.  Different variable selection methods were proposed for different models, such as linear model, logistic model and generalized linear model. However,  fewer works focused on variable selection for single index models, especially, for single index logistic model, due to the difficulty arose from the unknown link function and the slow mixing rate of MCMC algorithm for traditional logistic model. In this paper, we proposed a Bayesian variable selection procedure for single index logistic model by taking the advantage of Gaussian process and data augmentation. Numerical results from simulations and real data analysis show the advantage of our method over the state of arts.
\end{abstract}

\noindent%
\vfill

\newpage
\spacingset{1.5} 
\section{Introduction}
\label{sec:intro}
Single index regression model is defined by 
\begin{equation*}
	Y=m(\bm{X})+\epsilon=g(\bm{X}^T\bm{\beta})+\epsilon,
\end{equation*}
where $\bm{X} \in R^p$ is the covariate vector, the coefficient vector $\bm{\beta} \in\mathbb{R}^p$ is called single index, $g:\mathbb{R}\rightarrow\mathbb{R}$ is the unknown link function and $\epsilon$ is the noise term.  The constraint that $\|\bm{\beta}\|_2=1$ with the first non-zero element being positive is usually put to ensure the identifiability of $\bm{\beta}$. By assuming the response only relates to a single linear combination of the covariates,  single index model provides an efficient way for dimensional reduction in high-dimensional problems, which is often used in econometrics.

Based on observations $\{(y_i, \bm{x}_i): i=1, 2, \cdots, n)\}$, the interest is to estimate the single index parameter $\bm{\beta}$. The difficulty of this aim arises from the unknown link function $g$, and there are some different strategies to overcome the difficulty. Firstly, one may approximate the link function $g$ by base function expansion (e.g. splines, B-splines, kernel expansion)  or its local linear expansion \citep{hardle1993,Ichimura1993,xia2002,wang2009spline,radchenko2015, kuchibhotla2020,feng2020}, and then obtain its estimator $\hat{\bm{\beta}}$ as $\hat{\bm{\beta}}= \arg _{\bm{\beta}}  \min \sum_{i=1}^{n}\psi\left(y_{i}, \hat{g}(\bm{x}_{i}^{T}\bm{\beta})\right)$ where $\hat{g}$ is an estimator of the unknown link function. Secondly, the relationship that $m^{\prime}(\bm{X}) = \partial g(\bm{X}^{T}\bm{\beta})/\partial \bm{X} = g^{\prime}(\bm{X}^{T}\bm{\beta})\bm{\beta}$  leads to the average derivative estimator \citep{hardle1989,horowitz1996,hristache2001}.  Although the estimator is proved to have some nice property, additional parameters such as bandwidths are difficult to choose for real problems. In addition, Bayesian estimation methods for single index regression are proposed to estimate simultaneously the single index parameter and unknown link function. For example, by putting a Gaussian process prior on the unknown link function $g$, estimating $\bm{\beta}$ follows naturally into the Bayesian framework, see for example \citet{choi2011}.  However, when the sample size gets larger, it is very time-consuming to get the estimates. Differently, one may also approximate the unknown link function by B-splines and estimates all parameters in Bayesian way \citep{antoniadis2004, wang2009}. However, one should be careful in selecting parameters for B-splines. There are also other methods proposed for estimating parameters in single index models or extended single index models; see for example, \citet{cui2011,wang2015,li2017penalized}. 

Variable selection for single index model is more difficult than that for traditional models such as linear model and generalized linear model due to the unknown link function $g$.  Classical model selection methods such as AIC  \citep{naik2001}, cross validation \citep{kong2007} are extended to single index regression model, where the strategy is to use a nonparametric method to estimate $g$. Extending the penalized regression or penalized likelihood idea to single index model is straightforward by approximating the unknown link function $g$ using the basis function expansion; see for example, \citet{wang2008, wang2009, peng2011, zeng2012, radchenko2015, li2017, bindele2019}, or  by employing the sufficient dimension reduction methods; see for example, \citet{zhu2009, zhu2011, wang2015,lin2019}. 

Similar to the generalized linear model, it is natural to consider the generalized single index model by assuming the response $Y$ follows a distribution in the exponential family with probability density function $p(y|\theta, \phi) = \exp\{\frac{\theta y - b(\theta)}{a(\phi)}+c(y, \phi)\}$, where $a(\cdot)$, $b(\cdot)$ and $c(\cdot)$  are  measurable functions. Typically, we have $\mathbb{E}(Y)=\mu=b^\prime(\theta)$, and $\text{Var}(Y)=b^{\prime\prime}(\theta)a(\phi)$. Relating the covariates with the mean of $Y$ through the natural parameter $\theta$, i.e., $\theta=g(\bm{X}^T\bm{\beta})$, we get the generalized single index model.  There are some parameter estimation and variable selection methods proposed for this model; see, for example, \citet{cui2011, li2017}. Although these methods are proved to asymptotically have some nice properties, it is difficult to understand the impact of approximation error on the accuracy of variable selection in finite-sample settings. 

Compared with single index regression model, less attention is paid to the generalized single index model. Hence,  in this paper, we focus on the Single Index Logistic Model (SILM) defined as
\begin{equation}
\label{equ:logit}
Y\sim\text{Bernoulli}\left(\frac{\exp\{g(\bm{X}^{T}\bm{\beta})\}}{1+\exp\{g(\bm{X}^{T}\bm{\beta})\}}\right),
\end{equation}
and consider the parameter estimation and variable selection under this model. Specifically, we propose a Bayesian method with a Gaussian process prior for unknown link function $g$ and a spike-slab prior for single index $\bm{\beta}$ to achieve variable selection. Because the single index vector $\bm{\beta}$ is assumed to be on the unit sphere for the identifiability, a posterior projection method with data augmentation is applied to rapidly sample from the posterior distribution.

The rest of the paper is organized as follows. Section \ref{sec:meth} presents the Bayesian inference for SILM with details on priors for parameters, posterior distributions and posterior sampling algorithm. In Section \ref{sec:num}, we present numerical results from simulation studies and real-data analysis. Section \ref{sec:conc} concludes this paper with further discussions.

\section{Bayesian Inference for SILM}
\label{sec:meth}

\subsection{Prior Distribution}
\label{subsec:prior}
In Bayesian inference framework, the prior distribution summarizes the prior understanding of the statistical model.
In this section, we present prior distributions for unknown parameters in the single index logistic model.  Before that, we introduce some notations for convenience. Let $\left\{(y_i,\bm{x}_i): i=1,2,\cdots, n\right\}$ be $n$ independent and identically distributed (i.i.d) samples from SILM (\ref{equ:logit}), where $y_i \in \{-1, 1\}$ is the binary response variable, and $\bm{x}_i \in R^p$ is the $p$-dimensional predictor vector. Let $\bm{\beta}_0 \in R^p$ be the true value of the single index with $s$ non-zero elements, and $g_0$ be the true link function.
\subsubsection{Prior for Link Function}
Estimating the unknown link function $g$ is key to estimate the single index and variable selection in both single index regression model and logistic model. Approximation methods such as splines and B-splines are commonly employed in both frequentist and Bayesian literature, see for example, \citet{antoniadis2004,wang2009spline,wang2009,radchenko2015,lu2016,li2017,feng2020}. Although these methods enjoy computational advantage, one should be careful on selecting the order and knots of spline basis functions and an additional penalty on the number of knots is necessary to avoid over-fitting, which introduces some additional computational inefficiency. Gaussian process is popular  in Bayesian nonparametric modelling \citep{williams2006}, and has been successfully applied to single index regression model \citep{choi2011} and single index quantile regression model \citep{hu2013}, which motivates us put a Gaussian process prior on the unknown link function $g$.

Gaussian process is a stochastic process such
that any finite subcollection of random variables follows a multivariate Gaussian distribution. The function $f:\mathcal{X}\rightarrow\mathbb{R}$ follows Gaussian process prior, or the collection of random variables $\{f(x): x\in\mathcal{X}\}$ is said to be drawn from Gaussian process with mean function $m(\cdot)$ and covariance function $k(\cdot,\cdot)$, denoted by $$f(\cdot)\sim\text{GP}\left(m(\cdot),k(\cdot,\cdot)\right),$$ 
if for any set with finite elements $x=\{x_1,\cdots,x_n\}\subset \mathcal{X}$, the corresponding random vector $\{f(x_1), \cdots, f(x_n)\}$ follows a multivariate Gaussian distribution with mean vector $m(x)=\left(m(x_1),\cdots,m(x_n)\right)^T$ and covariance matrix $K(x)$, where $K_{ij}= k(x_{i}, x_{j})$.

A Gaussian process is completely specified by its mean and covariance function. Usually, the mean function is chosen to be identically zero. But the covariance function $k(\cdot,\cdot)$, also called kernel function, must be chosen such that the resulting covariance matrix $K$ is always positive semidefinite. For the single index logistic model, we assume the unknown link function $g$ follows a Gaussian process prior given by 
\begin{equation}
	\label{gpiror}
	g(\cdot) | \tau, l \sim \text{GP}(0,k(\cdot, \cdot)),
\end{equation}
with $k(\cdot, \cdot)$ defined by
\begin{equation}
\label{equ:ker}
k(\bm{x}_i, \bm{x}_j) = \tau\exp\{ -\frac{\left(\bm{x}_{i}^T\bm{\beta}-\bm{x}_{j}^T\bm{\beta}\right)^{2}}{l} \},
\end{equation}
where $\tau$ and $l$ are hyperparameters, and both assumed to have a inverse Gamma prior, that is, $$p(\tau)\propto\tau^{-(a_{\tau}+1)}\exp\{-\frac{b_{\tau}}{\tau}\}, \text{and}~ p(l)\propto l^{-(a_{l}+1)}\exp\{-\frac{b_{l}}{l}\}.$$

\subsubsection{Prior for Single Index}
In high-dimensional problems, it is often that only a small part of covariates truly contributes to the response, which requires that the estimated single index only has a small number of non-zero elements.  A classic framework to attain this requirement is to put a Spike-Slab prior (e.g., \citet{mitchell1988,george1993,ishwaran2005}) on the single index, that is, the prior of $\bm{\beta}=(\beta_1, \cdots, \beta_p)$ is given by 
\begin{equation}
p(\beta_j|\delta_j) = (1-\delta_j)p_{\text{Spike}}(\beta_j) + \delta_j p_{\text{Slab}}(\beta_j),j=1,2,\cdots,p,
\end{equation}  
where $\delta_j=1$ indicates the corresponding variable $X_j$ should be included in the model and $\delta_j=0$ indicates $X_j$ should not be included, $p_{\text{Slab}}(\beta_j)$ is a flat slab distribution (e.g. normal distribution) and $p_{\text{Spike}}(\beta_j)$ is a spike distribution (e.g. Dirac distribution) concentrating its mass to values close to zero.
To alleviate the sampling difficulty of posterior distribution,   \citet{george1993} proposed a hierarchical normal mixture spike-slab model, where the spike distribution is choose as a normal distribution with very small variance. 
Inspired by this, we propose our spike-slab prior for $\bm{\beta}$ as follows:
\begin{equation}\label{priorb1}
\beta_j|\delta_j,\sigma_{\beta_j}\sim\text{Normal}\left(0,r(\delta_j)\sigma_{\beta_j}\right), \sigma_{\beta_j}\sim\text{InvGamma}(a_{\sigma_{\beta_j}},b_{\sigma_{\beta_j}}),j=1,2,\cdots,p,
\end{equation}
where $\delta_j\in\{0,1\}$, $r(0)=c\ll 1$, $r(1)=1$, and  $\{a_{\sigma_{\beta_j}}, b_{\sigma_{\beta_j}}\}_{j=1}^{p}$ are hyperparameters. Integrating out $\sigma_{\beta_j}$, $\beta_j$ marginally follows Student distribution,
\begin{equation}\label{priorb2} \beta_j|\delta_j\sim\text{t}_{2a_{\sigma_{\beta_j}}}\left(0,\sqrt{r(\delta_j)b_{\sigma_{\beta_j}}/a_{\sigma_{\beta_j}}}\right).
\end{equation}
Finally, we put independent Bernoulli prior on the binary variables $\delta_j$, i.e.,
\begin{equation}\label{priorb3}
 \delta_j|\pi_j \sim \text{Bernoulli}(\pi_j),j=1,2,\cdots,p.
\end{equation}
The prior means variable $X_j$ is assumed to be included in the model with probability $\pi_i$. After incorporating the information from data, we include variable $X_j$ if the posterior probability of $\delta_j=1$ is larger than 0.5.

An important thing to be mentioned is that the single index $\bm{\beta}$ is constrained on the unit sphere $S^{p-1}$ for model identifiability. However, $\bm{\beta}$ from models (\ref{priorb1})-(\ref{priorb3}) is not assured to satisfy this constraint. Usually, priors like Uniform distribution on the unit sphere or half of the unit sphere and the Fisher-von Mises distribution are considered in the literature \citep{antoniadis2004,wang2009,choi2011}. However, the sparsity is ignored in these settings.
\citet{patra2018} developed a general approach, called posterior projection, for constrained Bayesian inference, which samples firstly from an unconstrained or less constrained posterior and then projects the draws to the constrained space through minimal distance mapping. We adopt this strategy for sampling $\bm{\beta}$ with our spike-slab prior.

\subsection{Posterior Distribution}
\label{subsec:post}
In Bayesian framework, the posterior distribution summarizes the information from prior knowledge (prior distribution) and data (likelihood).
Before presenting the main results, we introduce some notations to simplify the presentation.
Let $\bm{y}=\left(y_1,\cdots,y_n\right)$, $\bm{x}=(\bm{x}_1, \cdots, \bm{x}_n)$, $\bm{g}=\left(g_1,\cdots,g_n\right)$ with $g_i = g(\bm{x}_i^T\bm{\beta})$, $\bm{\delta}=\left(\delta_1,\cdots,\delta_p\right)$, $\bm{\sigma_\beta} = \left(\sigma_{\beta_1},\cdots,\sigma_{\beta_p}\right)$, and $\bm{\pi}=\left(\pi_1,\cdots,\pi_p\right)$. 
According to models (\ref{equ:logit})-(\ref{priorb3}), the posterior distribution of unknown parameters $\bm{g}$, $\bm{\beta}$, $\bm{\delta}$, and $\bm{\sigma_\beta}$, and hyperparameters $\bm{\pi},\tau$ and $l$, is given by 
\begin{equation}\label{poster}
p(\bm{g},\bm{\beta},\bm{\sigma_\beta},\bm{\delta},\bm{\pi},\tau,l|\bm{y}, \bm{x})\propto L(\bm{y} |\bm{g},\bm{\beta},\bm{\sigma_\beta},\bm{\delta},\bm{\pi},\tau,l, \bm{x})p(\bm{g},\bm{\beta},\bm{\sigma_\beta},\bm{\delta},\bm{\pi},\tau,l |\bm{x})
\end{equation}
where $L(\bm{y} |\bm{g},\bm{\beta},\bm{\sigma_\beta},\bm{\delta},\bm{\pi},\tau,l, \bm{x})$ is the likelihood function, 
$$p(\bm{g},\bm{\beta},\bm{\sigma_\beta},\bm{\delta},\bm{\pi},\tau,l |\bm{x}) = p(\bm{g}|\bm{\beta},\tau,l ,\bm{x})p(\bm{\beta}|\bm{\sigma_\beta},\bm{\delta})p(\bm{\sigma_\beta})p(\bm{\delta}|\bm{\pi})p(\bm{\pi})p(\tau)p(l),$$ and $p(\bm{\pi})$, $p(\tau)$ and $p(l)$ is the prior of $\bm{\pi},\tau$ and $l$, respectively.

According to the single index logistic model \ref{equ:logit}, we have 
\begin{equation} \label{likef}
L(\bm{y} |\bm{g},\bm{\beta},\bm{\sigma_\beta},\bm{\delta},\bm{\pi},\tau,l, \bm{x})=p(\bm{y}|\bm{g})=\prod_{i=1}^{n} \frac{\exp\left\{y_i g_i \right\}}{1+\exp\left\{ y_i g_i \right\}}.
\end{equation}
Moreover, according to model (\ref{gpiror}), we have
\begin{equation*}
p(\bm{g}|\bm{\beta},\tau,l,\bm{x})\propto|K|^{-1/2}\exp\{ -\frac{1}{2}\bm{g}^{T}K^{-1}\bm{g}\}.
\end{equation*}
In addition, according to models (\ref{priorb1})-(\ref{priorb3}), we have 
\begin{equation*}
p(\bm{\beta}|\bm{\sigma_\beta},\bm{\delta})\propto \prod_{j=1}^{p}\left(r(\delta_j)\sigma_{\beta_j}\right)^{-1/2}\exp\{-\frac{\beta_j^{2}}{2r(\delta_j)\sigma_{\beta_j}}\}\\
=|\Sigma_{\bm{\beta}}|^{-1/2}\exp\{ -\frac{1}{2}\bm{\beta}^{T}\Sigma_{\bm{\beta}}^{-1}\bm{\beta} \},
\end{equation*}
$$p(\bm{\sigma_\beta})\propto \prod_{j=1}^{p}\sigma_{\beta_j}^{-(a_{\sigma_{\beta_j}}+1)}\exp\{-\frac{b_{\sigma_{\beta_j}}}{\sigma_{\beta_j}}\},
\text{and}~ p(\bm{\delta}|\bm{\pi})=\prod_{j=1}^{p}\pi_j^{\delta_j}(1-\pi_j)^{1-\delta_j},$$
where $\Sigma_{\bm{\beta}} =\text{diag}\left(r(\delta_1)\sigma_{\beta_1},\cdots,r(\delta_p)\sigma_{\beta_p}\right)$.

\subsection{Posterior Sampling}
\subsubsection{Data Augmentation Scheme}
\label{subsec:aug}
Gibbs sampling algorithm is efficient for high-dimensional sampling problems by sampling from conditional marginal distributions. Usually, the Metropolis-Hasting (MH) algorithm \citep{hastings1970monte} is used to sample from the posterior distribution when the corresponding marginal distributions do not have a analytic form.  However, this MH within Gibbs algorithm is not so efficient for many problems due to a low accept rate.  Data augmentation is an efficient alternative for some problems to MH algorithm when the target distribution is not analytically available. For example, \citet{albert1993} proposed a data augmentation (DA)  approach for the probit model.  \citet{holmes2006} and \citet{polson2013} proposed DA methods for the logistic model. Following \citet{polson2013}, we have 
\begin{equation}
	\label{equ:pg}
	\begin{aligned}
		\frac{\left(\exp\{\psi\}\right)^a}{(1+\exp\{\psi\})^b}&=2^{-b}\exp\{\kappa\psi\}\mathbb{E}\left(\exp\{-\omega\psi^{2}/2\}\right) \\ &=2^{-b}\exp\{\kappa\psi\}\int_{0}^{\infty}\exp\{-\omega\psi^2/2\}p(\omega|b,0)d\omega,
	\end{aligned}
\end{equation}
where $b>0$, $a\in\mathbb{R}$, $\kappa=a-b/2$ and $p(\omega|b,0)$ is the density of $\text{P{\'o}lya-Gamma}(b,0)$ distribution, denoted as $\text{PG}(b,0)$. Generally, a $\text{PG}(b,c)$ random variable has the density
\begin{equation*}
	p(\omega|b,c) = \cosh^b\left(\frac{c}{2}\right)\exp\{-\frac{c^2}{2}\omega\}p(\omega|b,0).
\end{equation*}
Thus, we can rewrite the likelihood function (\ref{likef}) as
\begin{equation}
	\label{equ:pglogit}
		p(\bm{y}|\bm{g})=\prod_{i=1}^{n}\frac{\exp\{y_i g_i\}}{1+\exp\{y_i g_i\}}
		=\prod_{i=1}^{n}2^{-1}\exp\{y_i g_i/2\}\int_{0}^{\infty}\exp\{-\omega_i g_i^2/2\}p(\omega_i|1,0)d\omega_i,
\end{equation}
which suggests that the likelihood function is proportional to the marginal density function of $p(\bm{y},\bm{w}|\bm{g})= 2^{-n}\exp\{\frac{1}{2}\bm{y}^{T}\bm{g} - \frac{1}{2}\bm{g}^{T}\Omega \bm{g}\}\prod_{i=1}^{n}p(\omega_i|1,0)$,
where $\bm{\omega}=\left(\omega_1,\cdots,\omega_n\right)$, and $\Omega=\text{diag}(\omega_1,\cdots,\omega_n)$.

By adding the auxiliary variables $\bm{\omega}$ that follow the P{\'o}lya-Gamma distribution, we have,
\begin{equation}\label{equ:post}
\begin{aligned}
p(\bm{g},\bm{\beta}| \bm{\omega},\bm{\sigma_\beta},\bm{\delta},\bm{\pi},\tau,l, \bm{y}, \bm{x})\propto& p(\bm{y},\bm{\omega}|\bm{g})p(\bm{g},\bm{\beta}| \bm{\sigma_\beta},\bm{\delta},\bm{\pi},\tau,l,\bm{x})\\
\propto& \exp\{\frac{1}{2}\bm{y}^{T}\bm{g} - \frac{1}{2}\bm{g}^{T}\Omega \bm{g}\}\prod_{i=1}^{n}p(\omega_i|1,0)\\
&\times p(\bm{g}, \bm{\beta} | \bm{\sigma_\beta},\bm{\delta},\bm{\pi},\tau,l,\bm{x}).
\end{aligned}
\end{equation}
This simple data augmentation provides an efficient sampling algorithm. Details are given in next section.
\subsubsection{Sampling Algorithm}
In this section, we present the sampling algorithm that samples from the posterior distribution (\ref{poster}).  According to results presented in previous sections, we have following marginal distributions with detailed derivation given in Appendix \ref{app:cond}.
\begin{equation}
\label{equ:post:g}
\bm{g}|\bm{y},\bm{\omega},\bm{\beta},\bm{\sigma_\beta},\bm{\delta},\bm{\pi},\tau,l{\color{black},\bm{x}}\sim\text{Normal}\left(\bm{\mu_{g}},\Sigma_{g}\right),
\end{equation}
$$\omega_i|\bm{y},\bm{g},\bm{\beta},\bm{\sigma_\beta},\bm{\delta},\bm{\pi},\tau,l{\color{black},\bm{x}}\sim\text{PG}(1,g_i),i=1,2,\cdots,n,$$
\begin{equation}
\label{equ:post:beta}
p(\bm{\beta}|\bm{y},\bm{\omega},\bm{g},\bm{\sigma_\beta},\bm{\delta},\bm{\pi},\tau,l{\color{black},\bm{x}})\propto |K|^{-\frac{1}{2}}\exp\{ -\frac{1}{2}\bm{g}^TK^{-1}\bm{g} \}\times\exp\{ -\frac{1}{2}\bm{\beta}^{T}\Sigma_{\beta}^{-1}\bm{\beta} \},
\end{equation}
\begin{equation}
	\label{equ:post:delta}
	\delta_j|\bm{y},\bm{\omega},\bm{g},\bm{\beta},\bm{\sigma_\beta},\bm{\pi},\tau,l{\color{black},\bm{x}} \sim\text{Bernoulli}\left( \left(1+\frac{p_{\text{Spike}}(\beta_j)(1-\pi_j)}{p_{\text{Slab}}(\beta_j)\pi_j}\right)^{-1}\right),j=1,\cdots,p,
\end{equation}
\begin{equation}
	\label{equ:post:tau}
	\tau|\bm{y},\bm{\omega},\bm{g},\bm{\beta},\bm{\sigma_\beta},\bm{\delta},\bm{\pi},l{\color{black},\bm{x}}\sim \text{InvGamma}\left(a_\tau+\frac{n}{2},\frac{1}{2}\bm{g}^TK_0^{-1}\bm{g}+b_\tau\right),~K_0 = K/\tau,
\end{equation}
\begin{equation}
	\label{equ:post:l}
	p(l|\bm{y},\bm{\omega},\bm{g},\bm{\beta},\bm{\sigma_\beta},\bm{\delta},\bm{\pi},\tau{\color{black},\bm{x}})\propto |K|^{-\frac{1}{2}}\exp\{ -\frac{1}{2}\bm{g}^TK^{-1}\bm{g} \}\times\frac{1}{l^{a_l+1}}\exp\{-\frac{b_l}{l}\}.
\end{equation}
\begin{equation}
\sigma_{\beta_j}|\bm{y},\bm{\omega},\bm{g},\bm{\beta},\bm{\delta},\bm{\pi},\tau,l{\color{black},\bm{x}}\sim\text{InvGamma}\left(a_{\sigma_{\beta_j}}+\frac{1}{2},\frac{\beta_j^2}{2r(\delta_j)}+b_{\sigma_{\beta_j}}\right),j=1,\cdots,p,
\end{equation}
\begin{equation}
\pi_j|\bm{y},\bm{\omega},\bm{g},\bm{\beta},\bm{\sigma_\beta},\bm{\delta},\tau,l{\color{black},\bm{x}}\sim\text{Beta}\left(a_\pi+\delta_j,b_\pi+1-\delta_j\right),j=1,\cdots,p,
\end{equation}
where $\bm{\mu_g}=\frac{1}{2}\Sigma_g\bm{y}$, $\Sigma_g = K - K(K + \Omega^{-1})^{-1}K$,
$\Sigma_\beta = \text{diag}\left(r(\delta_1)\sigma_{\beta_1},\cdots,r(\delta_p)\sigma_{\beta_p}\right)$, and $p_{\text{Spike}}(\cdot)$ and $p_{\text{Slab}}(\cdot)$ represent the density functions for zero-mean Gaussian distributions with variance $c\cdot\sigma_{\beta_j}$ and $\sigma_{\beta_j}$ respectively.

It can be seen from equation (\ref{equ:post:beta}) and equation (\ref{equ:post:l}) that sampling  $\bm{\beta}$ and $l$ directly from their conditional posteriors is difficult, therefore a MH sampler is used. Note that the matrix $K$ is always ill-conditioned in practice, and sometimes $K$ may have a few small but negative eigenvalues due to the finite precision and round-off error, which causes problems to compute the inverse and determinant of matrix $K$ in equations (\ref{equ:post:beta}), (\ref{equ:post:tau}) and (\ref{equ:post:l}).  Partially collapsed Gibbs sampler \citep{van2008} is used to avoid this problem.  Specifically, by integrating out $\bm{g}$ from joint distribution of $(\bm{\beta},\bm{g})$, $(\tau,\bm{g})$ and $(l,\bm{g})$, we have following reduced conditional distributions for $\bm{\beta}, \tau$ and $l$.
\begin{equation}\label{equ:post:beta2}
\begin{aligned}
p(\bm{\beta}|\bm{y},\bm{\omega},\bm{\sigma_\beta},\bm{\delta},\bm{\pi},\tau,l{\color{black},\bm{x}})
=&\int p(\bm{\beta},\bm{g}|\bm{y},\bm{\omega},\bm{\sigma_\beta},\bm{\delta},\bm{\pi},\tau,l{\color{black},\bm{x}})d\bm{g}\\
\propto& |K+\Omega^{-1}|^{-\frac{1}{2}}\exp\{-\frac{1}{8}\bm{y}^{T}\Omega^{-1}(K+\Omega^{-1})^{-1}\Omega^{-1}\bm{y}\}\\
& \times \exp\{ -\frac{1}{2}\bm{\beta}^{T}\Sigma_{\beta}^{-1}\bm{\beta} \}.
\end{aligned}
\end{equation}
\begin{equation}\label{equ:post:tau2}
\begin{aligned}
p(\tau|\bm{y},\bm{\omega},\bm{\beta},\bm{\sigma_\beta},\bm{\delta},\bm{\pi},l{\color{black},\bm{x}})\propto&|K+\Omega^{-1}|^{-\frac{1}{2}}\exp\{-\frac{1}{8}\bm{y}^{T}\Omega^{-1}(K+\Omega^{-1})^{-1}\Omega^{-1}\bm{y}\}\\
& \times \frac{1}{\tau^{a_{\tau}+1}}\exp\{-\frac{b_{\tau}}{\tau}\},
\end{aligned}
\end{equation}
\begin{equation}\label{equ:post:l2}
\begin{aligned}
p(l|\bm{y},\bm{\omega},\bm{\beta},\bm{\sigma_\beta},\bm{\delta},\bm{\pi},\tau{\color{black},\bm{x}})\propto& |K+\Omega^{-1}|^{-\frac{1}{2}}\exp\{-\frac{1}{8}\bm{y}^{T}\Omega^{-1}(K+\Omega^{-1})^{-1}\Omega^{-1}\bm{y}\}\\
& \times  \frac{1}{l^{a_{l}+1}}\exp\{-\frac{b_{l}}{l}\}.
\end{aligned}
\end{equation}
Now, we need the inverse and determinant of $K+\Omega^{-1}$ instead of $K$, which is much more stable in practice. There is one thing to be mentioned, a improper sampling order may destroy the desired stationary distribution due to the partially collapse of $\bm{g}$ \citep{van2008,van2015}. According to suggestions in \citet{van2015}, we draw in order  $\bm{\beta},\tau,l,\bm{g},\bm{\omega},\bm{\sigma_\beta},\bm{\delta}$ and $\bm{\pi}$ iteratively, and the full sampling algorithm is presented in Algorithm \ref{algo:mhpcg}.

\begin{breakablealgorithm}
	\caption{Sampling Algorithm for SILM}
	\label{algo:mhpcg}
 	\begin{algorithmic}
	\REQUIRE Initial values $\{\bm{\beta}^{0},\tau^{0},l^{0},\bm{g}^{0},\bm{\omega}^{0},\bm{\sigma_\beta}^{0},\bm{\delta}^{0},\bm{\pi}^{0}\}$; Tuning parameters for proposal distributions $\{\sigma_\beta, \sigma_\tau,\sigma_l\}$; Spike parameter $r(0)=c$; Parameters for prior: $\{a_{\sigma_{\beta_j}},b_{\sigma_{\beta_j}}\}_{j=1}^{p}$, $\{a_\tau,b_\tau\}$, $\{a_l,b_l\}$, $\{a_\pi,b_\pi\}$; Data $\{\bm{x_i},y_i\}_{i=1}^{n}$ where $y_i\in\{1,-1\}$.
	
	\ENSURE Given $\bm{\beta}^{(t)},\tau^{(t)},l^{(t)},\bm{g}^{(t)},\bm{\omega}^{(t)},\bm{\sigma_\beta}^{(t)},\bm{\delta}^{(t)},\bm{\pi}^{(t)}$, generate $\bm{\beta}^{(t+1)},\tau^{(t+1)},l^{(t+1)},\bm{g}^{(t+1)},\bm{\omega}^{(t+1)}$, $\bm{\sigma_\beta}^{(t+1)},\bm{\delta}^{(t+1)},\bm{\pi}^{(t+1)}$ as follows:
	\begin{enumerate}
		\item Sample $\bm{\beta}^{(t+1)}$ from reduced conditional posterior $p(\bm{\beta}|\bm{y},\bm{\omega}^{(t)},\bm{\sigma_\beta}^{(t)},\bm{\delta}^{(t)},\bm{\pi}^{(t)},\tau^{(t)},l^{(t)})$ by Metropolis-Hastings (MH) algorithm with a normal proposal distribution $N(\bm{\beta}^{(t)},\sigma_\beta I_p)$, and $\bm{\beta}^{(t+1)}\leftarrow \bm{\beta}^{(t+1)}/||\bm{\beta}^{(t+1)}||_2$.
		
		\item Sample $\tau^{(t+1)}$ from reduced conditional posterior $p(\tau|\bm{y},\bm{\omega}^{(t)},\bm{\beta}^{(t+1)},\bm{\sigma_\beta}^{(t)},\bm{\delta}^{(t)},\bm{\pi}^{(t)},l^{(t)})$ by MH algorithm with proposal distribution $\ln \tilde{\tau}^{(t+1)}\sim\text{Normal}(\ln \tau^{(t)}, \sigma_\tau)$.
		
		\item Sample $l^{(t+1)}$ from reduced conditional posterior $p(l|\bm{y},\bm{\omega}^{(t)},\bm{\beta}^{(t+1)},\bm{\sigma_\beta}^{(t)},\bm{\delta}^{(t)},\bm{\pi}^{(t)},\tau^{(t+1)})$ by MH algorithm with proposal distribution $\ln \tilde{l}^{(t+1)}\sim\text{Normal}(\ln l^{(t)}, \sigma_l)$.
		
		\item Sample $\bm{g}^{(t+1)}$ from $\text{Normal}\left(\bm{\mu_{g}},\Sigma_{g}\right)$ with $\Sigma_g=K - K(K + \Omega^{-1})^{-1}K$, $\bm{\mu_g}=\frac{1}{2}\Sigma_g\bm{y}$, $\Omega =\text{diag}(\bm{\omega}^{(t)})$ and $K_{ij} = \tau^{(t+1)}\exp\{ -\frac{\left(\bm{x_{i}}^T\bm{\beta}^{(t+1)}-\bm{x_{j}}^T\bm{\beta}^{(t+1)}\right)^{2}}{l^{(t+1)}} \}$.
		
		\item Sample $\omega_i^{(t+1)}$ from PG$(1,g_i^{(t+1)})$ for $i = 1,2,\cdots,n$.
		
		\item Sample $\sigma_{\beta_j}^{(t+1)}$ from InvGamma$\left(a_{\sigma_{\beta_j}}+\frac{1}{2},\frac{\left(\beta_j^{(t+1)}\right)^2}{2r(\delta_j^{(t)})}+b_{\sigma_{\beta_j}}\right)$ for $j=1,2,\cdots,p$.
		
		\item Sample $\delta_j^{(t+1)}$ from Bernoulli$\left(\left(1+\frac{p_{\text{Spike}}(\beta_j^{(t+1)})(1-\pi_j^{(t)})}{p_{\text{Slab}}(\beta_j^{(t+1)})\pi_j^{(t)}}\right)^{-1}\right)$ for $j=1,2,\cdots,p$ that $p_{\text{Spike}}$ and $p_{\text{Slab}}$ represent the density functions for zero-mean Gaussian distributions with variance $c\cdot\sigma_{\beta_j}^{(t+1)}$ and $\sigma_{\beta_j}^{(t+1)}$ respectively.
		
		\item Sample $\pi_j^{(t+1)}$ from Beta$\left(a_\pi+\delta_j^{(t+1)},b_\pi+1-\delta_j^{(t+1)}\right)$ for $j=1,2,\cdots,p$.
	\end{enumerate}
\end{algorithmic}
\end{breakablealgorithm}

To make sure that the Markov chain is well mixed,  the first half of the $N$ MCMC samples are taken as burn-in period, and the second half of the MCMC samples are used for statistical inference. We take $N=10000$ in our analysis and the posterior mean are reported as the point estimation for each parameter of interest. In addition, Markov chains with independent random initial values are run to make sure the convergence of the Markov chain, and the Potential Scale Reduction Factors (PSRF) \citep{brooks1998} is used to diagnose the convergence. In practice, a PSRF less than 1.2 indicates the convergence of Markov chains \citep{brooks1998}.
\subsection{Computational  Issues}
\label{subsec:comput}

The major computational cost for our sampling algorithm comes from computing the determinant and inverse, or the linear equations of positive definite (PD) matrix $K+\Omega^{-1}\in\mathbb{R}^{n\times n}$ in equations (\ref{equ:post:g}), (\ref{equ:post:beta2}), (\ref{equ:post:tau2}) and (\ref{equ:post:l2}). Typically, Cholesky factorization is commonly used to compute the determinant and inverse of PD matrix, which takes $O(n^3)$ time for the computation. When the sample size goes to large, the computation burden is large. Thus, iterative methods, such as conjugate gradient method  with proper preconditioner, are suggested for this case.

In addition, approximation methods can be applied to speed up the computation of the inverse matrix. For example, Nystr\"{o}m approximation method is an important low-rank matrix approximation technique for the kernel matrix and has been widely developed \citep{williams2001,drineas2005,zhang2008,kumar2009,li2010,wang2013,gittens2016}. The Nystr\"{o}m method approximates $K$ by $\tilde{K}=CW^{\dag}C^{T}$, where $C\in\mathbb{R}^{n\times m}$ is consist of $m$ randomly selected columns from $K$, $W^{\dag}$ denotes the (pseudo) inverse of $W\in\mathbb{R}^{m\times m}$, which is the intersection of the selected columns and rows. By Woodbury-Sherman-Morrison identity, we have
\begin{equation}
\label{equ:nys}
\left(\tilde{K} + \Omega^{-1} \right)^{-1}= \Omega-\Omega C\left(W+C^T\Omega C\right)^{-1}C^T\Omega,
\end{equation}
and thus, we only need the inverse of matrix $W+C^T\Omega C\in\mathbb{R}^{m\times m}, m<n$.
Consequently, the total computational time for computing the inverse matrix can be reduced by Nystr\"{o}m method, for example, to $O(m^2n)$ time complexity as shown in \citet{williams2001}.

There is one thing needed to be mentioned that the approximation for posterior covariance $\Sigma_g$ in equation (\ref{equ:post:g}) may lead to indefinite matrix $\tilde{\Sigma}_g =K - K(\tilde{K} + \Omega^{-1})^{-1}K$. To ensure
the positive definiteness of covariance, we take the following positive semidefinite (PSD) approximation procedure:
$$\tilde{\Sigma}_g^+ = \arg\min_{A \geq 0} \|A - \tilde{\Sigma}_g\|_{F}^{2},$$
where $\|\cdot\|_F$ is Frobenius norm and $A \geq 0$ is symmetric PSD matrix with order $n$. Let $\tilde{\Sigma}_g = U\Lambda U^T$ be the eigen decomposition of $\tilde{\Sigma}_g$, and the optimal solution $\tilde{\Sigma}_g^+$ has analytical form $\tilde{\Sigma}_g^+ = U\Lambda^+ U^T$ where $\Lambda=\text{diag}\left(\lambda_1,\cdots,\lambda_n\right)$ and $\Lambda^+=\text{diag}\left(\max\{\lambda_1,0\},\cdots,\max\{\lambda_n,0\}\right)$ \citep{higham1988}. Even though the time complexity for eigen decomposition is still $O(n^3)$, but such decomposition is inevitable to sample from $n$-dimensional multivariate Gaussian distribution. 

Finally, several methods are proposed for approximating the determinant of a larger matrix, see for example, \citep{zhang2007,boutsidis2017}, \citep{han2015}, \citep{dong2017}, and \citet{gardner2018}, and can be used to speed up our MCMC algorithm.

\section{Numerical Results}
\label{sec:num}
In this section, we present results from simulation studies and real data analysis to evaluate the performance of the propose method.  For our sampling algorithm, parameters for proposal distributions are tuned such that the acceptance rates are around 20\% $\sim$ 30\%.  In addition, we take the spike parameter $c=r(0)=1/1000$, $a_{\sigma_{\beta_j}} =b_{\sigma_{\beta_j}}=0.5$ for $j=1, 2, \cdots, p$ and $a_\pi=b_\pi=a_l=b_l=a_{\tau}=b_{\tau}=0.5$. To show the advantages of the proposed method, we compare it with two state-of-art methods, the sparse sliced inverse regression method via Lasso (LassoSIR) \citep{lin2019} and distribution-based LASSO (DLASSO) method \citep{wang2015}. DLASSO has a tuning parameter  selected by the modified BIC criterion \citep{wang2015} with an extra parameter $\gamma$. Thus, we consider two different versions of DLASSO with $\gamma=0$ or $\gamma=0.5$, denoted by DLASSO$_0$ and DLASSO$_{0.5}$ respectively. Other settings of these two methods follow that given in the original paper.

\subsection{Simulations}
\label{subsec:simu}
Let $s$ be the number of variables truly contribute to the response, we consider the following three different models for the unknown like function $g$, and take the number of variables $p=10$. The link function $g$ in Model 1 is linear, which is the simplest one. In Model 2, $g$ is nonlinear and continuously differentiable. In Model 3, it is nonlinear and not differentiable at $t=0$, which adds some difficulty to the problem. 

\textbf{Model 1}: $g(t) = 5t$ with $\bm{\beta}^T = (3,2,2,0,0,0,0,0,0,0)/\sqrt{17} $ and $s=3$;

\textbf{Model 2}: $g(t) = 5\left(t + \sin(t^3)\right)$ with $\bm{\beta}^T=(2,2,1,1,0,0,0,0,0,0)/\sqrt{10}$ and $s=4$;

\textbf{Model 3}: $g(t) = 10|t|\times\sin(t)$ with $\bm{\beta}^T = (1,1,1,1,1,0,0,0,0,0)/\sqrt{5}$ and $s=5$;

In our simulations, the predictors $\{\bm{x}_i\}_{i=1}^{n}$ are generated from multivariate Gaussian distribution N$(0,\Sigma_X)$. We firstly take $\Sigma_X=\Sigma_1 = I_p$ to show the performance of different methods on the case where predictors are independent, and then take $\Sigma_X=\Sigma_2=\left(0.5^{|i-j|}\right)_{1\leq i,j\leq p}$ to see their performance on the case where predictors are dependent. The sample size $n$ is taken as $\{60, 100, 140\}$.  For parameter estimation, the estimation bias with the corresponding standard deviation is reported to evaluate the performance of each method. For variable selection, the number of true positives and false positives are reported to evaluate their performance. All results are reported based on 200 independent repetitions.

\begin{table}
	\centering
	\caption{TP/FP (standard deviation) from each method under different models.}
	\label{table:simtpfp}
	\resizebox{140mm}{78mm}
	{
		\begin{tabular}{cccccc}
		\toprule
			Method & $n$ & $\Sigma_X$ & Model 1 ($s=3$) & Model 2 ($s=4$) & Model 3 ($s=5$)\\
			\midrule
			Proposed&60&$\Sigma_1$&2.85(0.38)/0.12(0.35)&3.29(0.64)/0.09(0.28)&4.29(0.84)/0.07(0.25)\\
			&&$\Sigma_2$&2.61(0.56)/0.20(0.44)&2.95(0.68)/0.07(0.25)&3.67(0.77)/0.16(0.39)\\
			&100&$\Sigma_1$&3.00(0.00)/0.04(0.19)&3.76(0.47)/0.02(0.14)&4.95(0.21)/0.06(0.27)\\
			&&$\Sigma_2$&2.92(0.27)/0.06(0.23)&3.29(0.53)/0.12(0.38)&4.54(0.62)/0.03(0.17)\\
			&140&$\Sigma_1$&3.00(0.00)/0.01(0.10)&3.93(0.25)/0.00(0.00)&5.00(0.00)/0.00(0.00)\\
			&&$\Sigma_2$&2.95(0.21)/0.03(0.17)&3.57(0.53)/0.01(0.10)&4.90(0.30)/0.00(0.00)\\
			\hline
			 LassoSIR&60&$\Sigma_1$&2.99(0.10)/2.79(2.01)&3.90(0.30)/2.84(1.69)&4.97(0.17)/2.86(1.37)\\
			&&$\Sigma_2$&2.96(0.19)/1.79(1.62)&3.76(0.42)/1.87(1.64)&3.92(0.96)/1.52(1.43)\\
			&100&$\Sigma_1$&3.00(0.00)/3.11(1.87)&3.96(0.24)/2.72(1.83)&5.00(0.00)/3.06(1.31)\\
			&&$\Sigma_2$&3.00(0.00)/2.25(1.79)&3.92(0.27)/2.28(1.65)&4.30(0.83)/1.27(1.42)\\
			&140&$\Sigma_1$&3.00(0.00)/3.17(1.98)&4.00(0.00)/3.10(1.57)&5.00(0.00)/2.78(1.48)\\
			&&$\Sigma_2$&3.00(0.00)/1.93(2.04)&3.99(0.10)/2.18(1.71)&4.75(0.53)/1.64(1.28)\\
			\hline
			DLASSO$_0$&60&$\Sigma_1$&2.24(1.07)/0.12(0.35)&2.05(1.30)/0.06(0.23)&0.77(1.60)/0.01(0.10)\\
			&&$\Sigma_2$&2.33(0.82)/0.06(0.23)&2.27(1.15)/0.02(0.14)&0.77(1.25)/0.01(0.10)\\
			&100&$\Sigma_1$&2.81(0.56)/0.16(0.36)&3.35(0.88)/0.11(0.31)&3.02(2.32)/0.10(0.30)\\
			&&$\Sigma_2$&2.90(0.30)/0.06(0.31)&3.36(0.81)/0.03(0.17)&1.46(1.61)/0.01(0.10)\\
			&140&$\Sigma_1$&3.00(0.00)/0.11(0.31)&3.61(0.73)/0.08(0.27)&4.68(1.19)/0.23(0.44)\\
			&&$\Sigma_2$&2.98(0.14)/0.08(0.30)&3.64(0.59)/0.03(0.22)&2.73(1.88)/0.08(0.27)\\
			\hline
			
			DLASSO$_{0.5}$&60&$\Sigma_1$&1.07(1.30)/0.01(0.10)&0.86(1.18)/0.00(0.00)&0.01(0.50)/0.00(0.00)\\
			&&$\Sigma_2$&1.76(1.13)/0.02(0.14)&1.50(1.25)/0.00(0.00)&0.15(0.65)/0.00(0.00)\\
			&100&$\Sigma_1$&2.87(0.44)/0.09(0.32)&2.86(1.30)/0.03(0.17)&1.53(2.27)/0.04(0.19)\\
			&&$\Sigma_2$&2.77(0.58)/0.03(0.17)&2.96(1.10)/0.01(0.10)&0.73(1.21)/0.02(0.20)\\
			&140&$\Sigma_1$&3.00(0.00)/0.05(0.21)&3.56(0.85)/0.09(0.28)&4.64(1.28)/0.31(0.61)\\
			&&$\Sigma_2$&2.97(0.17)/0.08(0.44)&3.48(0.65)/0.03(0.17)&1.38(1.71)/0.05(0.26)\\
			\bottomrule
		\end{tabular}
	}
\end{table}

Table \ref{table:simtpfp} shows the simulation results on variable selection from each method under different models. When sample size is small ($n=60$), the performance of {\color{black} LassoSIR} is slightly better than the proposed method in terms of a little higher TP, but the corresponding FP is relatively more higher than our method. When sample size is larger, say $n>100$, the proposed method almost has the best performance among these methods in terms of both TP and FP. The performance of DLASSO is the worst among them. Importantly,  the dependence between predictors has a strong impact on its performance under Model 3. This suggests that DLASSO may provide unsatisfied results in some cases.  For the proposed method, the impact of dependence between predictors is minor.

\begin{table}
	\centering
	\caption{Biases and standard error (S.E.) of estimators from the proposed method under different models.} \label{table:simbeta}
	\resizebox{130mm}{105mm}
	{
		\begin{tabular}{ccccccccccccc}
			\toprule
			Model & $n$ & & $\beta_1$ & $\beta_2$& $\beta_3$& $\beta_4$& $\beta_5$& $\beta_6$& $\beta_7$& $\beta_8$& $\beta_9$& $\beta_{10}$\\ \midrule
			\textbf{\textit{Model 1}}& \\
			\multirow{6}{*}{$\Sigma_X=\Sigma_1$}
			& \multirow{2}{*}{60} & Bias &-.030&-.052&-.048&-.001&.005&-.005&-.006&-.004&-.006&-.001 \\
			& & S.E. &.086&.135&.118&.046&.050&.067&.066&.057&.077&.053 \\
			& \multirow{2}{*}{100} & Bias 
			&-.030&-.002&-.009&.001&.001&.001&-.001&.002&-.001 &-.008 \\
			& & S.E. &.052&.060&.065&.046&.038&.041&.045&.045&.043&.041\\
			& \multirow{2}{*}{140} & Bias &.052&.060&.065&.046&.038&.041&.045&.045&.043&.041\\
			& & S.E. &.038&.050&.057&.033&.036&.032&.038&.037&.032&.041\\
			\midrule
			\multirow{6}{*}{$\Sigma_X=\Sigma_2$}
			& \multirow{2}{*}{60} & Bias 
			&-.077&-.087&-.066&.028&.006&-.003&-.005&.010&-.002&-.001 \\
			& & S.E. &.157&.199&.159&.116&.071&.058&.078&.068&.069&.085\\
			& \multirow{2}{*}{100} & Bias 
			&-.046&-.033&-.020&.005&.007&.008&-.003&.001&.001&.002 \\
			& & S.E. &.079&.130&.123&.063&.047&.056&.044&.039&.041&.049\\
			& \multirow{2}{*}{140} & Bias &-.024&-.029&-.025&-.001&-.001&.005&-.005&-.001&.001&-.004\\
			& & S.E. &.066&.121&.103&.059&.035&.045&.042&.039&.037&.039\\
			\midrule
			\textbf{\textit{Model 2}}& \\
			\multirow{6}{*}{$\Sigma_X=\Sigma_1$}
			& \multirow{2}{*}{60} & Bias 
			&-.025&-.031&-.062&-.066&-.006&.005&-.001&.006&.005&-.011\\
			& & S.E. &.108&.081&.127&.120&.054&.058&.065&.062&.058&.049\\
			& \multirow{2}{*}{100} & Bias 
			&-.025&-.013&-.012&-.015&.002&-.002&.003&.003&-.006&.002\\
			& & S.E. &.061&.057&.081&.086&.036&.041&.035&.035&.050&.041\\
			& \multirow{2}{*}{140} & Bias 
			&-.014&-.008&-.010&-.012&-.009&.001&.000&-.001&-.004&.001\\
			& & S.E. &.046&.046&.061&.073&.032&.037&.029&.034&.034&.029\\
			\midrule
			\multirow{6}{*}{$\Sigma_X=\Sigma_2$}
			& \multirow{2}{*}{60} & Bias 
			&-.096&-.042&-.061&-.075&.004&.003&.006&.007&.002&.001\\
			& & S.E. &.153&.158&.188&.170&.056&.053&.057&.058&.061&.066\\
			& \multirow{2}{*}{100} & Bias 
			&-.039&-.028&-.046&-.050&.029&.008&.004&-.001&-.005&.002\\
			& & S.E. &.092&.099&.139&.128&.082&.055&.049&.056&.050&.049\\
			& \multirow{2}{*}{140} & Bias 
			&-.018&-.033&-.033&-.016&.009&-.001&-.003&-.003&-.001&-.002\\
			& & S.E. &.078&.098&.115&.100&.033&.041&.040&.040&.034&.036\\
			\midrule
			\textbf{\textit{Model 3}}& \\
			\multirow{6}{*}{$\Sigma_X=\Sigma_1$}
			& \multirow{2}{*}{60} & Bias 
			&-.063&-.072&-.063&-.052&-.057&.004&.005&.000&.002&.001\\
			& & S.E. &.148&.144&.137&.162&.152&.052&.065&.064&.055&.064\\
			& \multirow{2}{*}{100} & Bias 
			&-.025&-.014&-.024&-.018&-.022&-.000&.002&.003&-.004&.001\\
			& & S.E. &.082&.068&.069&.088&.091&.039&.051&.040&.046&.048\\
			& \multirow{2}{*}{140} & Bias 
			&-.019&-.013&.002&-.018&-.016&-.001&.002&-.001&-.001&-.002\\
			& & S.E. &.065&.058&.062&.068&.061&.032&.042&.036&.031&.036\\
			\midrule
			\multirow{6}{*}{$\Sigma_X=\Sigma_2$}
			& \multirow{2}{*}{60} & Bias
			&-.122&-.092&-.025&-.062&-.131&.033&.001&.013&.006&-0.002\\
			& & S.E. &.167&.194&.189&.184&.183&.084&.090&.073&.065&0.061\\
			& \multirow{2}{*}{100} & Bias 
			&-.038&-.038&-.041&-.037&-.056&.009&.000&.002&.002&-.004\\
			& & S.E. &.107&.139&.133&.150&.124&.047&.039&.043&.047&.044\\
			& \multirow{2}{*}{140} & Bias 
			&-.030&-.017&-.023&-.029&-.020&.003&-.003&.002&-.001&.003\\
			& & S.E. &.086&.104&.103&.092&.086&.036&.038&.040&.030&.033\\
			\bottomrule
		\end{tabular}
	}
\end{table}

Furthermore, we show in Table \ref{table:simbeta} the estimation bias of parameters from the proposed method, and the corresponding results for LassoSIR and DLASSO are shown in Appendix~\ref{app:result}. Results in Table \ref{table:simbeta}  indicate that our proposed method estimates the single index $\beta$ well in all cases, and the biases and standard errors decline in general as the sample sizes increase. The bias from LassoSIR and DLASSO are larger than that from our method, which may partially explain the better performance of our method on variable selection.

\begin{table}
	\centering
	\caption{TP/FP (standard deviation) from the proposed method under Model 1 when Nystr\"{o}m approximation is used to speed up calculation.}\label{table:simtpfp2}
	\begin{tabular}{ccc}
		\toprule
		$m$ & $\Sigma_X=\Sigma_1$ & $\Sigma_X=\Sigma_2$ \\
		\midrule
		$60$&2.96(0.19)/0.58(0.92)&2.80(0.42)/1.47(1.31)\\
		$80$&2.99(0.10)/0.09(0.37)&2.93(0.29)/0.37(0.78)\\
		$100$&3.00(0.00)/0.00(0.00)&2.99(0.10)/0.07(0.35)\\ \bottomrule
	\end{tabular}
\end{table}

\subsection{Accelerating MCMC for Large Sample Problems}
\label{subsec:acce}

In this section, we evaluate the impact of the approximation in computing the inverse and determinant of positive definite matrix $K+\Omega^{-1}\in\mathbb{R}^{n\times n}$ in equations (\ref{equ:post:g}), (\ref{equ:post:beta2}), (\ref{equ:post:tau2}) and (\ref{equ:post:l2}), on variable selection performance of the proposed method.  Since the proposed method has similar performance on different link functions, we now focus on Model 1, and take sample size $n=500$.

We focus on the performance of the Nystr\"{o}m approximation method as described in Section~\ref{subsec:comput}. Specifically, we follow the method in \citet{williams2001} to approximate the inverse of matrix $K+\Omega^{-1}$, where $m \in \{60, 80, 100\}$ columns are chosen uniformly at random without replacement, and the determinant is approximated using GPytorch \citep{gardner2018}. Theoretically the matrix $W+C^T\Omega C$ in equation (\ref{equ:nys}) is at least positive semidefinite, but numerically it has a few small but negative eigenvalues like $K$. To tackle with the problem, jilters are added to the diagonal of $K$, i.e. $K$ is replaced by $K+\epsilon I_{n}$. In practice, $\epsilon= 1/100\cdot\tau$ and the inverse of $W+C^T\Omega C$ is computed by Cholesky factorization. Results given in Table~\ref{table:simtpfp2} show that the proposed method with Nystr\"{o}m approximation has a comparable performance with that with exact calculation as shown in Table~\ref{table:simtpfp}. This shows the hope of speeding up the MCMC algorithm for problems with a large sample size. There is one thing to be mentioned that the convergence of MCMC algorithm is fast and one thousand iterations are enough for obtaining converged MCMC samples.

\subsection{Real Data Examples}
\label{subsec:real}
In this section, we show the performance of the proposed method on three real examples. To eliminate the impact of units, predictor variables in all examples are scaled to have mean zero and standard deviation one.  We report the estimated $\beta$, $\bm{\hat{\beta}}$, from different methods. For the proposed method, we also report the standard deviation of $\bm{\hat{\beta}}$ (denoted as $s_{\bm{\hat{\beta}}}$), the posterior probability of $\delta_j=1$ (denoted as $\hat{\pi_j}$),  standard deviation of $\hat{\pi_j}$ (denoted as $s_{\hat{\pi_j}}$), and PRSF for convergence diagnosis of Markov chains.

\textbf{Example 1}: Swiss Banknote Data \citep{flury1988}.  This dataset, available in the R package uskewFactors, contains measurements on 200 Swiss banknotes including 100 genuine and 100 counterfeit. There are 6 predictor variables that are measurement results on the banknotes including length of bill, width of left edge, width of right edge, bottom margin width, top margin width and diagonal length. For convenience, these variables are denoted by $X_1,\cdots,X_6$ respectively. The response $Y$ is coded as -1 or 1, indicating whether the banknote is genuine or not. 

\begin{table}
	\centering
	\caption{ Results for Swiss Banknote Data (\textbf{Example 1})}
	\label{table:banknote}
		\begin{tabular}{lcccccc}
			\toprule
			Method&$X_1$&$X_2$&$X_3$&$X_4$&$X_5$&$X_6$\\
			\midrule
			LassoSIR&0&-0.282&0.306&0.451&0.474&-0.630\\
			DLASSO$_0$&0&0&0&0.638&0.376&-0.671\\
			DLASSO$_{0.5}$&0&0&0&0.638&0.376&-0.671\\
			\midrule
			\multicolumn{7}{l}{\bf{\textit{Proposed}}}\\
			$\bm{\hat{\beta}}$&-0.001&0.031&0.029&\textbf{0.667}&\textbf{0.299}&\textbf{-0.603}\\
			$s_{\bm{\hat{\beta}}}$ &0.057&0.101&0.098&0.103&0.197&0.162\\
			$\hat{\pi_j}$&0.107&0.151&0.158&\textbf{0.888}&\textbf{0.592}&\textbf{0.861}\\
			$s_{\hat{\pi_j}}$&0.310&0.358&0.365&0.315&0.491&0.345\\
			PRSF&1.03&1.05&1.05&1.05&1.11&1.10\\
			\bottomrule
		\end{tabular}
\end{table}
 Table \ref{table:banknote} shows the estimation and variable selection results from different methods. For the propose method, 6 Markov chains are run from different initials and PSRFs for each $\beta_j$ are also shown in Table \ref{table:banknote}, which indicates Markov chain is converged.  The proposed method and DLASSO select $\{X_4, X_5, X_6\}$ as the true variables contribute to the response in this example, however LassoSIR selects two more variables, $X_2$ and $X_3$.

\textbf{Example 2}: Body Fat Data \citep{penrose1985}. This dataset, available in the R package mfp, contains 252 observations and 128 samples are left after deleting samples with errors in the R package mplot. The response in this dataset is the percentage of body fat determined by underwater weighing and the predictor variables include age, weight, height and ten body circumference measurements (neck, chest, abdomen, hip, thigh, knee, ankle, biceps, forearm and wrist). For convenience, these variables are denoted by $X_1,\cdots,X_{13}$, respectively.

This dataset has been widely studied based on single index regression model, see for example, \citet{li2017,peng2011} and \citet{li2017penalized}. In our study, to fit to our settings, the response variable is transformed to be binary depending on whether its value is greater than the sample median, indicating whether the man is overweight or not. Table \ref{table:fat} shows the estimation and variable selection results from different methods. It can be seen that the predictor $X_6$ (abdomen) is selected by the proposed method and DLASSO, which coincides with the empirical fact that the measurement of abdomen indeed generally reflect the body fat. However, LassoSIR selects much more predictors among which the measurements on chest $(X_5)$ and abdomen $(X_6)$ is the most influential.

\begin{table}
	\centering
	\caption{Results for Body Fat Data  (\textbf{Example 2})}
	\label{table:fat}
	\resizebox{160mm}{29mm}
	{
		\begin{tabular}{cccccccccccccc}
			\toprule
			Method&$X_1$&$X_2$&$X_3$&$X_4$&$X_5$&$X_6$&$X_7$&$X_8$&$X_9$&$X_{10}$&$X_{11}$&$X_{12}$&$X_{13}$\\
			\midrule
			LassoSIR &0.028&0.124&0&0.011&0.661&0.512&0.331&0&0.182&0&0&0.374&0\\
			DLASSO$_0$&0&0&0&0&0&1&0&0&0&0&0&0&0\\
			DLASSO$_{0.5}$&0&0&0&0&0&1&0&0&0&0&0&0&0\\
			\midrule
			\multicolumn{7}{l}{\bf{\textit{Proposed}}}\\ 	$\bm{\hat{\beta}}$&0.014&-0.009&0.015&-0.023&0.019&\textbf{0.965}&-0.025&-0.022&-0.074&-0.007&-0.026&-0.016&-0.048\\
			$s_{\bm{\hat{\beta}}}$&0.042&0.058&0.045&0.047&0.115&0.037&0.067&0.058&0.106&0.054&0.059&0.052&0.064\\
			$\hat{\pi_j}$&0.070&0.100&0.085&0.091&0.179&\textbf{0.879}&0.128&0.124&0.220&0.099&0.111&0.098&0.150\\
			$s_{\hat{\pi_j}}$&0.255&0.301&0.280&0.288&0.384&0.325&0.334&0.330&0.414&0.299&0.314&0.298&0.357\\
			PRSF&1.01&1.05&1.03&1.05&1.09&1.17&1.08&1.10&1.05&1.07&1.02&1.11&1.02\\
			\bottomrule
		\end{tabular}
	}
\end{table}

\textbf{Example 3}: Breast Cancer Data \citep{patricio2018}. This dataset,  available in the UCI machine learning repository, is developed to screen potential biomarkers of breast cancer based on routine blood analysis. Clinical features are observed or measured for 64 patients with breast cancer and 52 healthy controls. There are 9 predictors including Age, BMI, Glucose, Insulin, HOMA, Leptin, Adiponectin, Resistin and MCP-1, denoted by $X_1, \cdots, X_9$ respectively. Based on logistic regression, random forests and support vector machine methods, Age, BMI, Glucose and Resistin ($X_1,X_2,X_3,X_8$) are selected as potential biomarkers in \citet{patricio2018}.

\begin{table}
	\centering
	\caption{Results for Breast Cancer Data  (\textbf{Example 3})}
	\label{table:breast}
		\begin{tabular}{cccccccccc}
			\toprule
			Method&$X_1$&$X_2$&$X_3$&$X_4$&$X_5$&$X_6$&$X_7$&$X_8$&$X_9$\\
			\midrule
			LassoSIR&0.045&0.822&-0.455&-0.240&0&0&0&-0.237&0\\
			DLASSO$_0$&0&0&1&0&0&0&0&0&0\\
			DLASSO$_{0.5}$&0&0&1&0&0&0&0&0&0\\
			\midrule
		\multicolumn{7}{l}{\bf{\textit{Proposed}}}\\  $\bm{\hat{\beta}}$&0.053&\textbf{0.216}&\textbf{-0.698}&-0.131&-0.041&0.023&0.017&\textbf{-0.569}&0.011\\
			$s_{\bm{\hat{\beta}}}$&0.084&0.138&0.140&0.149&0.094&0.085&0.078&0.159&0.052\\
			$\hat{\pi_j}$&0.171&\textbf{0.539}&\textbf{0.888}&0.320&0.167&0.164&0.131&\textbf{0.849}&0.085\\
			$s_{\hat{\pi_j}}$&0.376&0.498&0.314&0.466&0.373&0.371&0.337&0.358&0.279\\
			PRSF&1.03&1.07&1.15&1.07&1.10&1.08&1.02&1.14&1.06\\
			\bottomrule
		\end{tabular}
\end{table}

Table~\ref{table:breast} shows the estimation and variable selection results from different methods. The results show that DLASSO methods only select predictor $X_3$ (Glucose). Comparing with results in \citet{patricio2018}, $X_1$ (Age) is not included in our selected model and $X_4$ (Insulin) is extra variable selected by LassoSIR. Thus, results from our method and LassoSIR are more closer to the underlying truth in \citet{patricio2018}.

\section{Conclusion}
\label{sec:conc}

In this paper, we propose a Bayesian estimation and variable selection approach for single index Logistic model with a relatively efficient MCMC sampling algorithm. Specially, the unknown link function $g$ is assumed to have a Gaussian process prior, and a Spike-Slab prior is assumed for single index $\bm{\beta}$ to find variables truly contributing to the response. The advantage of Bayesian model for single index Logistic model over frequentist methods is its ability to simultaneously estimate the unknown link function and the single index.  Numerical results from both simulation studies and real data analysis show the advantage of the proposed method.

Although we focus on parameter estimation and variable selection approach for single index Logistic model, the framework can be extended to other single index general models, for example, single index Poisson model. The problem for Bayesian estimation for single index models is the efficiency of MCMC algorithm. As discussed before, when the sample size is large, it will be very time consuming for sampling from the posterior distribution. Although some approximation methods (Section \ref{subsec:comput}) can be used to accelerate the MCMC algorithm and are shown to work well through simulations, the impact of this approximation on the distribution of MCMC samples are not well studied yet. It is in a strong necessity to explore the theoretical properties on the MCMC samples from algorithms with approximations. That is, we should study the convergence rate of MCMC samples from an approximated MCMC algorithms to the target distributions.
 In addition, for highly imbalanced problems, current sampling algorithm will be inefficient \citep{johndrow2019}. Thus, it is also in need to further investigation on how to tackle with the problem when samples are highly imbalanced. These interesting topics are left as future works.

\newpage
\bibliographystyle{chicago}
\bibliography{ref}

\appendix

\section{Conditional Posterior Distribution}
\label{app:cond}

After adding the auxiliary variables $\bm{\omega}$ (Section \ref{subsec:aug}), the posterior distribution for SILM is given by:
\begin{equation*}
\begin{aligned}
p(\bm{\omega},\bm{g},\bm{\beta},\bm{\sigma_\beta},\bm{\delta},\bm{\pi},\tau,l|\bm{y}, \bm{x})\propto& p(\bm{y},\bm{\omega}|\bm{g})p(\bm{g},\bm{\beta},\bm{\sigma_\beta},\bm{\delta},\bm{\pi},\tau,l{\color{black}|\bm{x}})\\
\propto& \exp\{\frac{1}{2}\bm{y}^{T}\bm{g} - \frac{1}{2}\bm{g}^{T}\Omega \bm{g}\}\prod_{i=1}^{n}p(\omega_i|1,0)\times p(\bm{g},\bm{\beta},\bm{\sigma_\beta},\bm{\delta},\bm{\pi},\tau,l{\color{black}|\bm{x}})\\
\propto& \exp\{\frac{1}{2}\bm{y}^{T}\bm{g} - \frac{1}{2}\bm{g}^{T}\Omega \bm{g}\}\prod_{i=1}^{n}p(\omega_i|1,0)\\
&\times |K|^{-1/2}\exp\{ -\frac{1}{2}\bm{g}^{T}K^{-1}\bm{g}\}\times\prod_{j=1}^{p}\left(r(\delta_j)\sigma_{\beta_j}\right)^{-1/2}\exp\{-\frac{\beta_j^{2}}{2r(\delta_j)\sigma_{\beta_j}}\}\\
&\times \prod_{j=1}^{p}\sigma_{\beta_j}^{-(a_{\sigma_{\beta_j}}+1)}\exp\{-\frac{b_{\sigma_{\beta_j}}}{\sigma_{\beta_j}}\}\times\prod_{j=1}^{p}\pi_j^{\delta_j}(1-\pi_j)^{1-\delta_j}\\
&\times \tau^{-(a_{\tau}+1)}\exp\{-\frac{b_{\tau}}{\tau}\}\times l^{-(a_{l}+1)}\exp\{-\frac{b_{l}}{l}\}\times\prod_{j=1}^{p}\pi_j^{a_{\pi}-1}(1-\pi_j)^{b_{\pi}-1}.
\end{aligned}
\end{equation*}

The conditional posteriors for $\left(\bm{\omega},\bm{g},\bm{\beta},\bm{\sigma_\beta},\bm{\delta},\bm{\pi},\tau,l\right)$ are as follows:
\begin{enumerate}
	\item $\bm{g}|\bm{y},\bm{\omega},\bm{\beta},\bm{\sigma_\beta},\bm{\delta},\bm{\pi},\tau,l , \bm{x}$
	\begin{equation*}
	\begin{aligned}
	p(\bm{g}|\bm{y},\bm{\omega},\bm{\beta},\bm{\sigma_\beta},\bm{\delta},\bm{\pi},\tau,l, \bm{x}) &\propto \exp\{\frac{1}{2}\bm{y}^{T}\bm{g} - \frac{1}{2}\bm{g}^{T}\Omega \bm{g}\}\times \exp\{ -\frac{1}{2}\bm{g}^{T}K^{-1}\bm{g} \}\\
	& \propto \exp\{-\frac{1}{2}\left(\bm{g}-\bm{\mu_g}\right)^T\Sigma_g^{-1}\left(\bm{g}-\bm{\mu_g}\right)\},
	\end{aligned}
	\end{equation*}
	with mean vector $\bm{\mu_g}=\frac{1}{2}\Sigma_g\bm{y}$ and covariance matrix $$\Sigma_g = (K\Omega+I_n)^{-1}K = K - K(K + \Omega^{-1})^{-1}K.$$
	
	\item $\bm{\omega}|\bm{y},\bm{g},\bm{\beta},\bm{\sigma_\beta},\bm{\delta},\bm{\pi},\tau,l, \bm{x}$
	\begin{equation*}
	\begin{aligned}
	p(\bm{\omega}|\bm{y},\bm{g},\bm{\beta},\bm{\sigma_\beta},\bm{\delta},\bm{\pi},\tau,l , \bm{x})\propto&\exp\{- \frac{1}{2}\bm{g}^{T}\Omega \bm{g}\}\prod_{i=1}^{n}p(\omega_i|1,0)\\
	=& \prod_{i=1}^{n}\exp\{- \frac{1}{2}g_i^2\omega_i\}p(\omega_i|1,0),
	\end{aligned}
	\end{equation*}
	Therefore we have $\omega_i|\bm{y},\bm{g},\bm{\beta},\bm{\sigma_\beta},\bm{\delta},\bm{\pi},\tau,l , \bm{x}\sim\text{PG}(1,g_i)$ for $i=1,2,\cdots,n$.
	
	\item $\bm{\beta}|\bm{y},\bm{\omega},\bm{g},\bm{\sigma_\beta},\bm{\delta},\bm{\pi},\tau,l , \bm{x}$	
	$$p(\bm{\beta}|\bm{y},\bm{\omega},\bm{g},\bm{\sigma_\beta},\bm{\delta},\bm{\pi},\tau,l, \bm{x})\propto |K|^{-1/2}\exp\{ -\frac{1}{2}\bm{g}^{T}K^{-1}\bm{g} \}\times\exp\{ -\frac{1}{2}\bm{\beta}^{T}\Sigma_{\beta}^{-1}\bm{\beta} \},  $$
	with $\Sigma_{\beta} = \text{diag}\left(r(\delta_1)\sigma_{\beta_1},\cdots,r(\delta_p)\sigma_{\beta_p}\right)$.

	\item $\bm{\sigma_{\beta}}|\bm{y},\bm{\omega},\bm{g},\bm{\beta},\bm{\delta},\bm{\pi},\tau,l , \bm{x}$
	\begin{equation*}
	\begin{aligned}
	p(\bm{\sigma_{\beta}}|\bm{y},\bm{\omega},\bm{g},\bm{\beta},\bm{\delta},\bm{\pi},\tau,l, \bm{x})\propto& 
	|\Sigma_\beta|^{-1/2}\exp\{ -\frac{1}{2}\bm{\beta}^{T}\Sigma_{\beta}^{-1}\bm{\beta} \}\times \prod_{j=1}^{p}\sigma_{\beta_j}^{-(a_{\sigma_{\beta_j}}+1)}\exp\{-\frac{b_{\sigma_{\beta_j}}}{\sigma_{\beta_j}}\}\\
	\propto&\prod_{j=1}^{p}\sigma_{\beta_j}^{-(a_{\sigma_{\beta_j}}+1/2+1)}\exp\left\{-\left( \frac{\beta_j^{2}}{2r(\delta_j)}+b_{\sigma_{\beta_j}}\right)/\sigma_{\beta_j}\right\},
	\end{aligned}
	\end{equation*}
	Therefore we have $\sigma_{\beta_j}|\bm{y},\bm{\omega},\bm{g},\bm{\beta},\bm{\delta},\bm{\pi},\tau,l , \bm{x}\sim\text{InvGamma}\left(a_{\sigma_{\beta_j}}+\frac{1}{2},\frac{\beta_j^2}{2r(\delta_j)}+b_{\sigma_{\beta_j}}\right)$ for $j=1,\cdots,p$.

	\item $\bm{\delta}|\bm{y},\bm{\omega},\bm{g},\bm{\beta},\bm{\sigma_\beta},\bm{\pi},\tau,l , \bm{x}$
	$$p(\bm{\delta}|\bm{y},\bm{\omega},\bm{g},\bm{\beta},\bm{\sigma_\beta},\bm{\pi},\tau,l, \bm{x})\propto\prod_{j=1}^{p}\left(r(\delta_j)\sigma_{\beta_j}\right)^{-1/2}\exp\{-\frac{\beta_j^{2}}{2r(\delta_j)\sigma_{\beta_j}}\}\pi_j^{\delta_j}(1-\pi_j)^{1-\delta_j},$$
	and for each $j=1,2,\cdots,p,$ we have
	$$\frac{p(\delta_j=1|\bm{y},\bm{\omega},\bm{g},\bm{\beta},\bm{\sigma_\beta},\bm{\pi},\tau,l, \bm{x})}{p(\delta_j=0|\bm{y},\bm{\omega},\bm{g},\bm{\beta},\bm{\sigma_\beta},\bm{\pi},\tau,l, \bm{x})} = \frac{\sigma_{\beta_j}^{-1/2}\exp\{-\beta_j^2/2\sigma_{\beta_j}\}\pi_j}{\left(r(0)\sigma_{\beta_j}\right)^{-1/2}\exp\{-\beta_j^2/2r(0)\sigma_{\beta_j}\}(1-\pi_j)}.$$
	Denote the density functions for zero-mean Gaussian distributions with variance $r(0)\cdot\sigma_{\beta_j}$ and $\sigma_{\beta_j}$ by $p_{\text{Spike}}(\cdot)$ and $p_{\text{Slab}}(\cdot)$ respectively, and then we have
	$$\delta_j|\bm{y},\bm{\omega},\bm{g},\bm{\beta},\bm{\sigma_\beta},\bm{\pi},\tau,l , \bm{x} \sim\text{Bernoulli}\left(\left(1+\frac{p_{\text{Spike}}(\beta_j)(1-\pi_j)}{p_{\text{Slab}}(\beta_j)\pi_j}\right)^{-1}\right),j=1,\cdots,p.$$

	\item $\bm{\pi}|\bm{y},\bm{\omega},\bm{g},\bm{\beta},\bm{\sigma_\beta},\bm{\delta},\tau,l , \bm{x}$
	$$p(\bm{\pi}|\bm{y},\bm{\omega},\bm{g},\bm{\beta},\bm{\sigma_\beta},\bm{\delta},\tau,l, \bm{x})\propto \prod_{j=1}^{p}\pi_j^{\delta_j}(1-\pi_j)^{1-\delta_j}\cdot\pi_j^{a_{\pi}-1}(1-\pi_j)^{b_{\pi}-1},$$
	hence $\pi_j|\bm{y},\bm{\omega},\bm{g},\bm{\beta},\bm{\sigma_\beta},\bm{\delta},\tau,l\sim\text{Beta}\left(a_\pi+\delta_j,b_\pi+1-\delta_j\right)$ for $j=1,\cdots,p$.
	
	\item $\tau|\bm{y},\bm{\omega},\bm{g},\bm{\beta},\bm{\sigma_\beta},\bm{\delta},\bm{\pi},l , \bm{x}$
	
	Let $K_0 = K/\tau$, i.e., $K_{0(i,j)} = \exp\{ -\frac{\left(\bm{x_{i}}^T\bm{\beta}-\bm{x_{j}}^T\bm{\beta}\right)^{2}}{l} \}$, then 
	\begin{equation*}
	\begin{aligned}
	p(\tau|\bm{y},\bm{\omega},\bm{g},\bm{\beta},\bm{\sigma_\beta},\bm{\delta},\bm{\pi},l, \bm{x})&\propto|K|^{-1/2}\exp\{ -\frac{1}{2}\bm{g}^{T}K^{-1}\bm{g} \} \times \tau^{-(a_{\tau}+1)}\exp\{-\frac{b_{\tau}}{\tau}\}\\
	&=\tau^{-n/2}|K_0|^{-1/2}\exp\{ -\frac{\bm{g}^{T}K_0^{-1}\bm{g}}{2\tau} \} \times \tau^{-(a_{\tau}+1)}\exp\{-\frac{b_{\tau}}{\tau}\}\\
	&=\tau^{-(a_{\tau}+1+n/2)}\exp\left\{-\left(\frac{1}{2}\bm{g}^{T}K_0^{-1}\bm{g}+b_{\tau}\right) /\tau \right\},
	\end{aligned}
	\end{equation*}
	hence $\tau|\bm{y},\bm{\omega},\bm{g},\bm{\beta},\bm{\sigma_\beta},\bm{\delta},\bm{\pi},l\sim \text{InvGamma}\left(a_\tau+\frac{n}{2},\frac{1}{2}\bm{g}^TK_0^{-1}\bm{g}+b_\tau\right)$.
	
	\item $l|\bm{y},\bm{\omega},\bm{g},\bm{\beta},\bm{\sigma_\beta},\bm{\delta},\bm{\pi},\tau , \bm{x}$
	$$p(l|\bm{y},\bm{\omega},\bm{g},\bm{\beta},\bm{\sigma_\beta},\bm{\delta},\bm{\pi},\tau, \bm{x})\propto |K|^{-\frac{1}{2}}\exp\{ -\frac{1}{2}\bm{g}^TK^{-1}\bm{g} \}\times\frac{1}{l^{a_l+1}}\exp\{-\frac{b_l}{l}\}.$$
\end{enumerate}

\section{Additional Results}
\label{app:result}

\begin{table}
	\centering
	\caption{Biases and standard errors (S.E.) of estimators from LassoSIR under different models.}
	\resizebox{130mm}{105mm}
	{
		\begin{tabular}{ccccccccccccc}
			\toprule
			Model & $n$ & & $\beta_1$ & $\beta_2$& $\beta_3$& $\beta_4$& $\beta_5$& $\beta_6$& $\beta_7$& $\beta_8$& $\beta_9$& $\beta_{10}$\\ 
			\midrule
		   \textbf{\textit{Model 1}}& \\
			\multirow{6}{*}{$\Sigma_X=\Sigma_1$}
			& \multirow{2}{*}{60} & Bias &-.012&-.022&-.051&.005&-.008&-.012&-.001&-.007&.007&-.006\\
			& & S.E. &.093&.124&.130&.085&.071&.082&.070&.089&.088&.081\\
			& \multirow{2}{*}{100} & Bias 
			&-.004&-.018&-.022&.001&.006&-.007&-.012&-.001&.004&.002\\
			& & S.E. &.075&.092&.087&.070&.059&.044&.072&.043&.059&.061\\
			& \multirow{2}{*}{140} & Bias &.011&-.024&-.027&-.004&.001&.001&-.003&-.009&.001&.000\\
			& & S.E. &.056&.076&.068&.055&.050&.044&.047&.052&.059&.045\\
			\midrule
			\multirow{6}{*}{$\Sigma_X=\Sigma_2$}
			& \multirow{2}{*}{60} & Bias 
			&-.042&-.040&-.057&.019&-.006&.015&.002&.013&-.018&.010\\
			& & S.E. &.160&.201&.195&.087&.059&.068&.083&.083&.091&.084\\
			& \multirow{2}{*}{100} & Bias 
			&-.020&-.005&-.062&.018&-.004&.003&-.002&-.001&.009&-.006\\
			& & S.E.&.112&.161&.134&.081&.057&.082&.064&.069&.066&.071\\
			& \multirow{2}{*}{140} & Bias &.001&-.015&-.048&.010&.011&-.011&.013&-.003&.011&-.007\\
			& & S.E.&.076&.125&.106&.060&.056&.062&.066&.062&.058&.046\\
						\midrule
			\textbf{\textit{Model 2}}& \\
			\multirow{6}{*}{$\Sigma_X=\Sigma_1$}
			& \multirow{2}{*}{60} & Bias 
			&.013&-.034&-.058&-.046&-.006&-.013&-.006&-.002&-.003&.010\\
			& & S.E. &.093&.089&.127&.138&.081&.065&.066&.078&.090&.070\\
			& \multirow{2}{*}{100} & Bias 
			&-.008&.004&-.048&-.038&-.001&-.006&-.008&.006&-.001&.007\\
			& & S.E. &.081&.081&.103&.100&.057&.072&.062&.059&.056&.053\\
			& \multirow{2}{*}{140} & Bias 
			&-.005&-.001&-.024&-.023&.006&.003&-.002&-.002&-.004&.004\\
			& & S.E.&.067&.063&.079&.085&.043&.051&.046&.044&.053&.055\\
			\midrule
			\multirow{6}{*}{$\Sigma_X=\Sigma_2$}
			& \multirow{2}{*}{60} & Bias 
			&-.028&-.049&-.020&-.107&.007&.012&.001&.014&.001&.001\\
			& & S.E. &.172&.194&.180&.152&.088&.067&.082&.088&.083&.091\\
			& \multirow{2}{*}{100} & Bias 
			&-.056&.004&-.010&-.081&.016&.014&-.001&.003&-.001&-.012\\
			& & S.E. &.132&.138&.148&.140&.089&.074&.080&.081&.075&.066\\
			& \multirow{2}{*}{140} & Bias 
			&-.027&-.011&-.007&-.046&.012&-.003&.012&-0.005&.001&.004\\
			& & S.E. &.100&.116&.141&.118&.074&.073&.075&.041&.045&.050\\
			\midrule
			\textbf{\textit{Model 3}}& \\
			\multirow{6}{*}{$\Sigma_X=\Sigma_1$}
			& \multirow{2}{*}{60} & Bias 
			&-.045&-.049&-.017&-.013&-.025&.001&.015&-.012&.009&.010\\
			& & S.E. &.112&.146&.125&.121&.121&.106&.103&.094&.112&.082\\
			& \multirow{2}{*}{100} & Bias 
			&-.010&-.011&-.022&-.015&-.018&.004&-.001&-.006&.008&-.003\\
			& & S.E. &.086&.104&.091&.092&.091&.078&.076&.070&.067&.071\\
			& \multirow{2}{*}{140} & Bias 
			&-.018&-.010&-.014&.004&-.018&.000&.012&-.008&.001&-.005\\
			& & S.E. &.079&.071&.079&.074&.082&.066&.064&.062&.064&.060\\
			\midrule
			\multirow{6}{*}{$\Sigma_X=\Sigma_2$}
			& \multirow{2}{*}{60} & Bias
			&-.178&-.227&-.167&-.192&-.230&.016&.038&-.006&-.003&.003\\
			& & S.E. &.244&.413&.371&.366&.302&.175&.133&.137&.118&.136\\
			& \multirow{2}{*}{100} & Bias 
			&-.148&-.160&-.136&-.147&-.236&.010&.005&.009&.008&-.017\\
			& & S.E. &.234&.382&.352&.356&.302&.112&.114&.126&.089&.097\\
			& \multirow{2}{*}{140} & Bias 
			&-.082&-.079&-.065&-.053&-.151&.038&-.001&.0207&.012&-.002\\
			& & S.E. &.190&.266&.264&.254&.218&.130&.110&.102&.090&.084\\
			\bottomrule
		\end{tabular}
	}
\end{table}

\begin{table}
	\centering
	\caption{Biases and standard errors (S.E.) of estimators from DLASSO$_{0}$ under different models.}
	\resizebox{130mm}{105mm}
	{
		\begin{tabular}{ccccccccccccc}
			\toprule
			Model & $n$ & & $\beta_1$ & $\beta_2$& $\beta_3$& $\beta_4$& $\beta_5$& $\beta_6$& $\beta_7$& $\beta_8$& $\beta_9$& $\beta_{10}$\\
		\midrule
		\textbf{\textit{Model 1}}& \\
			\multirow{6}{*}{$\Sigma_X=\Sigma_1$}
			& \multirow{2}{*}{60} & Bias &-.058&-.120&-.159&.003&.001&.001&.001&-.001&.001&-.002\\
			& & S.E. &.267&.262&.243&.030&.023&.029&.013&.035&.035&.025 \\
			& \multirow{2}{*}{100} & Bias 
			&.010&-.040&-.051&.005&.001&.001&-.001&-.001&.001&.000 \\
			& & S.E. &.101&.159&.162&.031&.032&.042&.029&.042&.026&.000\\
			& \multirow{2}{*}{140} & Bias &-.003&.002&-.011&-.001&-.001&-.003&.000&.001&.000&-.001\\
			& & S.E. &.053&.065&.068&.013&.025&.025&.000&.027&.000&.029\\
			\midrule
			\multirow{6}{*}{$\Sigma_X=\Sigma_2$}
			& \multirow{2}{*}{60} & Bias 
			&-.115&-.019&-.137&.019&.000&.000&.000&.000&.000&.000\\
			& & S.E. &.266&.284&.285&.078&.000&.000&.000&.000&.000&.000\\
			& \multirow{2}{*}{100} & Bias 
			&-.017&.000&-.045&-.001&.005&.000&-.004&-.002&.000&.000 \\
			& & S.E.&.090&.171&.157&.031&.039&.000&.046&.029&.000&.000\\
			& \multirow{2}{*}{140} & Bias &-.011&-.012&-.018&.007&.004&.002&.000&.003&.000&.000\\
			& & S.E.&.092&.132&.128&.038&.030&.021&.000&.032&.000&.000\\
			\midrule
			\textbf{\textit{Model 2}}& \\
			\multirow{6}{*}{$\Sigma_X=\Sigma_1$}
			& \multirow{2}{*}{60} & Bias 
			&-.141&-.096&-.222&-.199&.001&.000&-.005&.000&.003&-.001\\
			& & S.E. &.320&.315&.173&.183&.015&.000&.059&.000&.032&.016\\
			& \multirow{2}{*}{100} & Bias 
			&.003&-.013&-.073&-.086&.001&.001&-.002&.001&.000&-.001\\
			& & S.E. &.114&.137&.164&.169&.023&.016&.032&.038&.000&.018\\
			& \multirow{2}{*}{140} & Bias 
			&.002&-.001&-.070&-.030&.000&.004&.001&.000&.004&-.001\\
			& & S.E.&.073&.095&.144&.136&.000&.025&.023&.000&.031&.011\\
			\midrule
			\multirow{6}{*}{$\Sigma_X=\Sigma_2$}
			& \multirow{2}{*}{60} & Bias 
			&-.209&-.054&-.049&-.221&.000&.003&.000&.000&.000&.003\\
			& & S.E. &.300&.316&.283&.192&.000&.030&.000&.000&.000&.038\\
			& \multirow{2}{*}{100} & Bias 
			&-.046&-.030&-.030&-.056&.003&.000&.000&.000&-0.001&.000\\
			& & S.E. &.172&.182&.208&.197&.025&.000&.000&.000&.015&.000\\
			& \multirow{2}{*}{140} & Bias 
			&-.035&-.003&.026&-.081&.001&.000&.000&.000&.001&.000\\
			& & S.E. &.121&.105&.149&.164&.016&.000&.000&.000&.038&.000\\
				\midrule
			\textbf{\textit{Model 3}}& \\
			\multirow{6}{*}{$\Sigma_X=\Sigma_1$}
			& \multirow{2}{*}{60} & Bias 
			&-.348&-.356&-.386&-.380&-.358&-.001&.000&.000&.000&.000\\
			& & S.E. &.237&.216&.161&.172&.221&.017&.000&.000&.000&.000\\
			& \multirow{2}{*}{100} & Bias 
			&-.152&-.164&-.170&-.186&-.176&.001&-.003&-.005&-.001&-.002\\
			& & S.E. &.250&.236&.240&.229&.235&.025&.034&.028&.010&.024\\
			& \multirow{2}{*}{140} & Bias 
			&-.043&-.039&-.027&-.029&-.031&-.001&-.006&.009&-.001&-.004\\
			& & S.E. &.137&.120&.125&.124&.129&.045&.030&.048&.023&.024\\
			\midrule
			\multirow{6}{*}{$\Sigma_X=\Sigma_2$}
			& \multirow{2}{*}{60} & Bias
			&-.357&-.330&-.341&-.336&-.384&.002&.000&.000&.000&.000\\
			& & S.E. &.230&.271&.264&.258&.200&.024&.000&.000&.000&.000\\
			& \multirow{2}{*}{100} & Bias 
			&-.346&-.257&-.214&-.213&-.355&.000&0.000&.003&.000&.000\\
			& & S.E. &.202&.291&.343&.333&.210&.000&.000&.038&.000&.000\\
			& \multirow{2}{*}{140} & Bias 
			&-.265&-.102&-.151&-.153&-.203&.010&.006&.002&-.002&.000\\
			& & S.E. &.238&.306&.270&.277&.282&.051&.049&.029&.029&.000\\
			\bottomrule
		\end{tabular}
	}
\end{table}

\begin{table}
	\centering
	\caption{Biases and standard errors (S.E.) of estimators from DLASSO$_{0.5}$ under different models.}
	\resizebox{130mm}{105mm}
	{
		\begin{tabular}{ccccccccccccc}
			\toprule
			Model & $n$ & & $\beta_1$ & $\beta_2$& $\beta_3$& $\beta_4$& $\beta_5$& $\beta_6$& $\beta_7$& $\beta_8$& $\beta_9$& $\beta_{10}$\\
			\midrule
		\textbf{\textit{Model 1}}& \\
			\multirow{6}{*}{$\Sigma_X=\Sigma_1$}
			& \multirow{2}{*}{60} & Bias &-.381&-.320&-.322&-.003&.000&.000&.000&.000&.000&.000 \\
			& & S.E. &.401&.247&.258&.032&.000&.000&.000&.000&.000&.000 \\
			& \multirow{2}{*}{100} & Bias 
			&.004&-.020&-.041&-.002&.001&-.003&.000&.003&-.001&.000 \\
			& & S.E. &.081&.121&.159&.024&.015&.034&.000&.045&.019&.000\\
			& \multirow{2}{*}{140} & Bias &-.007&.006&-.008&-.001&-.003&.000&.000&.000&-.002&.000\\
			& & S.E. &.051&.059&.071&.014&.031&.000&.018&.000&.022&.000\\
			\midrule
			\multirow{6}{*}{$\Sigma_X=\Sigma_2$}
			& \multirow{2}{*}{60} & Bias 
			&-.242&-.080&-.256&.002&.000&.000&-.003&.000&.000&.000 \\
			& & S.E. &.365&.359&.278&.021&.000&.000&.036&.000&.000&.000\\
			& \multirow{2}{*}{100} & Bias 
			&-.036&-.017&-.058&.002&.002&.000&.000&.003&.000&.000 \\
			& & S.E.&.145&.200&.195&.028&.025&.000&.000&.039&.000&.000\\
			& \multirow{2}{*}{140} & Bias &-.020&-.008&-.006&.003&.001&.003&.000&-.002&.004&.000\\
			& & S.E.&.076&.137&.128&.026&.012&.027&.000&.029&.033&.000\\
				\midrule
			\textbf{\textit{Model 2}}& \\
			\multirow{6}{*}{$\Sigma_X=\Sigma_1$}
			& \multirow{2}{*}{60} & Bias 
			&-.383&-.376&-.293&-.273&.000&.000&.000&.000&.000&.000\\
			& & S.E. &.357&.362&.092&.124&.000&.000&.000&.000&.000&.000\\
			& \multirow{2}{*}{100} & Bias 
			&-.049&-.056&-.135&-.128&.000&.001&-.001&.001&.000&.000\\
			& & S.E. &.212&.227&.180&.178&.000&.016&.019&.014&.000&.000\\
			& \multirow{2}{*}{140} & Bias 
			&-.016&-.007&-.046&-.055&.000&.001&.003&.001&.003&.000\\
			& & S.E.&.121&.112&.135&.148&.000&.023&.027&.013&.024&.000\\
			\midrule
			\multirow{6}{*}{$\Sigma_X=\Sigma_2$}
			& \multirow{2}{*}{60} & Bias 
			&-.288&-.165&-.187&-.276&.000&.000&.000&.000&.000&.000\\
			& & S.E. &.351&.390&.234&.131&.000&.000&.000&.000&.000&.000\\
			& \multirow{2}{*}{100} & Bias 
			&-.104&-.001&-.044&-.142&.000&.000&.000&.000&.000&.003\\
			& & S.E. &.233&.198&.215&.199&.000&.000&.000&.000&.000&.030\\
			& \multirow{2}{*}{140} & Bias 
			&-.019&-.011&-.006&-.093&.003&.000&.000&.000&.000&.002\\
			& & S.E. &.108&.118&.181&.185&.026&.000&.000&.000&.000&.026\\
				\midrule
			\textbf{\textit{Model 3}}& \\
			\multirow{6}{*}{$\Sigma_X=\Sigma_1$}
			& \multirow{2}{*}{60} & Bias 
			&-.435&-.436&-.437&-.426&-.433&.000&.000&.000&.000&.000\\
			& & S.E. &.083&.078&.100&.127&.100&.000&.000&.000&.000&.000\\
			& \multirow{2}{*}{100} & Bias 
			&-.312&-.300&-.320&-.316&-.298&-.001&.001&.000&-.001&.000\\
			& & S.E. &.210&.219&.203&.205&.224&.019&.012&.000&.023&.000\\
			& \multirow{2}{*}{140} & Bias 
			&-.044&-.038&-.036&-.037&-.036&-.006&-.001&-.003&.009&-.002\\
			& & S.E. &.139&.132&.132&.133&.135&.037&.038&.027&.045&.035\\
			\midrule
			\multirow{6}{*}{$\Sigma_X=\Sigma_2$}
			& \multirow{2}{*}{60} & Bias
			&-.435&-.413&-.423&-.431&-.435&.000&.000&.000&.000&.000\\
			& & S.E. &.082&.155&.145&.094&.085&.000&.000&.000&.000&.000\\
			& \multirow{2}{*}{100} & Bias 
			&-.403&-.318&-.309&-.328&-.403&.002&.000&.003&.000&.000\\
			& & S.E. &.152&.279&.301&.265&.156&.024&.000&.031&.000&.000\\
			& \multirow{2}{*}{140} & Bias 
			&-.359&-.269&-.241&-.288&-.312&.008&.000&.001&.002&.000\\
			& & S.E. &.199&.289&.308&.266&.255&.0498&.000&.015&.020&.000\\
			\bottomrule
		\end{tabular}
	}
\end{table}

\end{document}